\pdfoutput=1

\documentclass{raa}
\usepackage{graphicx,times}
\usepackage{natbib}
\usepackage{amssymb,amsmath}

\usepackage{subfigure}
\usepackage{url}
\usepackage{CJK}
\usepackage{lscape}

\renewcommand{\vec}[1]{ {\mathbf #1} }

\newcommand{\Fig}{{Figure}}

\bibpunct{(}{)}{;}{a}{}{,}

\begin{document}

\title{A Comparison Study of a Solar Active-Region Eruptive Filament
  and a Neighboring Non-Eruptive Filament}

\author{Chaowei Jiang$^{1,2*}$, S. T. Wu$^{2,3}$, Xueshang~Feng$^{1}$, Qiang Hu$^{2,4}$}

\institute{$^{1}$SIGMA Weather Group, State Key Laboratory for Space
  Weather, National Space Science Center, Chinese
  Academy of Sciences, Beijing 100190\\
  $^{2}$Center for Space Plasma \& Aeronomic Research,
  $^{3}$Department of Mechanical \& Aerospace Engineering,
  $^{4}$Department of Space Sciences,
  The University of Alabama in Huntsville, Huntsville, AL 35899\\
  $^*$The corresponding author email: cwjiang@spaceweather.ac.cn\\
\vs \no
   %{\small accepted by RAA}
}

\abstract{Solar active region (AR) 11283 is a very magnetically
  complex region and it has produced many eruptions. However, there
  exists a non-eruptive filament in the plage region just next to an
  eruptive one in the AR, which gives us an opportunity to perform a
  comparison analysis of these two filaments. The coronal magnetic
  field extrapolated using a CESE--MHD--NLFFF code
  \citep{Jiang2013NLFFF} reveals that two magnetic flux ropes (MFRs)
  exist in the same extrapolation box supporting these two filaments,
  respectively. Analysis of the magnetic field shows that the eruptive
  MFR contains a bald-patch separatrix surface (BPSS) co-spatial very
  well with a pre-eruptive EUV sigmoid, which is consistent with the
  BPSS model for coronal sigmoids. The magnetic dips of the
  non-eruptive MFRs match H$\alpha$ observation of the non-eruptive
  filament strikingly well, which strongly supports the MFR-dip model
  for filaments.  Compared with the non-eruptive MFR/filament (with a
  length of about 200~Mm), the eruptive MFR/filament is much smaller
  (with a length of about 20~Mm), but it contains most of the magnetic
  free energy in the extrapolation box and holds a much higher free
  energy density than the non-eruptive one.  Both the MFRs are weakly
  twisted and cannot trigger kink instability. The AR eruptive MFR is
  unstable because its axis reaches above a critical height for torus
  instability, at which the overlying closed arcades can no longer
  confine the MFR stably.  On the contrary, the quiescent MFR is very
  firmly held by its overlying field, as its axis apex is far below
  the torus-instability threshold height. Overall, this comparison
  investigation supports that MFR can exist prior to eruption and the
  ideal MHD instability can trigger MFR eruption.  \keywords{Magnetic
    fields --- Sun: corona --- Sun: Filaments --- Sun: Eruptions} }

   \authorrunning{C.-W. Jiang et al. }            %author_head in even pages
   \titlerunning{Comparison Study of Two Solar Filaments}  % title_head in odd pages
   \maketitle

%________________________________________________ sections below
%

\section{Introduction}
\label{sec:1}

As a leading cause of space weather, coronal mass ejections (CMEs) are
closely correlated with solar filament eruptions. A recent statistic
study shows that more than 70 percent of filaments eventually erupt
and result in CMEs \citep{McCauley2015}. Therefore, it is of great
importance for space weather forecasting to understand why the
filaments erupt and predict when they are likely to erupt. To answer
these questions, one needs the information of a key parameter, the
invisible coronal magnetic field behind the filaments, which is
believed to play the primary role in supporting the filaments and
characterizing their stability.

Accordingly, a variety of theoretical models have been proposed to
explain the initiation of filament eruptions with the coronal magnetic
disruption as the basis \citep[see, e.g.,][and references
therein]{Forbes2006, Aulanier2010, Schmieder2013, Aulanier2014}.  The
physical mechanisms causing the eruption can be classified as, e.g.,
the breakout model \citep{Antiochos1999}, the tether cutting model
\citep{Moore2001} as well as ideal MHD instabilities \citep[e.g., kink
instability and torus instability,][]{Hood1981, Velli1990, Torok2004,
  Torok2005, Kliem2006} and the non-ideal MHD instabilities
\citep[e.g. tearing mode instability given by][]{Wu2000}.

It is understood that the pre-eruption, stressed core field in the
corona is usually kept to be stable by an envelope/overlying field
anchored strongly at the photosphere, thus, the essences of all
eruption initiation models are to find a way how the balance between
the outward magnetic pressure of the core field and the downward
magnetic tension of the overlying field can be broken. Hence,
\citet{Antiochos1999} proposed the breakout model basing on a
quadrupolar magnetic field configuration, and that the eruption is
triggered by reconnection at a coronal null above the sheared core
arcade, which removes the overlying flux and then allows the core to
escape, similar to a streamer and flux rope model \citep{WuGuo1997,
  Wu1997}. The tether cutting model usually refers to a complete
process of the formation and eruption of a magnetic flux rope (MFR)
from an initial simple, sheared bipolar magnetic arcade
\citep{Moore1992, Moore2001}.  In a strongly sheared bipolar field,
current sheet forms above the photospheric polarity inversion line
(PIL) and results in tether-cutting reconnection between the arcade
field lines, which progressively transforms the arcade core into a
rising twisted MFR. The reconnection is further forced by converging
motions toward the PIL, flux diffusion and cancellation in the
photosphere \citep{Amari2003A, Amari2003B, Linker2003}.  Such
processes continuously add flux to the MFR and in turn weaken the
photospheric anchorage below the MFR, until at some instant, the
remaining arcade tethers are too weak to prevent the eruption of the
MFR. In this case, it is different from the breakout model since the
magnetic reconnection works at low altitude below the stressed
field. The ideal MHD instabilities mainly include the kink instability
for a MFR which occurs, if the twist, a measure of the number of
windings of the field lines around the rope axis, exceeds a critical
value, leading to a helical deformation of the MFR's axis
\citep{Torok2005, Torok2010} and the torus instability
\citep{Kliem2006}, which is equivalent to a catastrophic loss of
equilibrium \citep{Forbes1990,Forbes1991,Kliem2014}, occurring when
the MFR reaches an unstable threshold determined by decay speed of the
envelope/overlying magnetic field that stabilizes the MFR. On the
other hand, the non-ideal MHD instability was performed by
\citet{Wu2000}, who have investigated the formation of observed plasma
blobs \citep{Sheeley1997} with multi-polarity magnetic field topology
due to tearing mode instability.

Of all the mentioned theoretical models, it appears that only the
ideal MHD-instability one can provide a set of specific parameters,
i.e., the twist degree of the MFR and/or the decay index of the
envelope field, that quantifies the condition when the magnetic
configuration might erupt. Thus this is a very attracting advantage
that will promisingly make itself most popular in the community of
space weather research. We note that such model is built up on a basic
element, the MFR, which is assumed to exist prior to the
eruption. This is indeed supported by several lines of evidence. The
coronal sigmoids often appearing in CME-productive ARs
\citep{Rust1996, Hudson1998, Canfield1999} indicate sheared and
twisted magnetic configurations, thus MFR, prior to the eruption;
Furthermore, the MFR can explain well the supporting of filaments
because the dip structure in MFR provides a reservoir where the cool
dense material can be held against gravity \citep{Mackay2010, Guo2010,
  Su2011}, and thus long-lasting quiescent filament strongly suggests
a correspondingly lasting MFR; On the other hand, with different
numerical techniques and force-free models, many authors have
successfully reconstructed from vector magnetogram MFRs that are
consistent with the observation signatures like sigmoids and filaments
\citep{Canou2010, Cheng2010, Guo2010, Jing2010, GuoY2013,
  Jiang2014formation, Jiang2014NLFFF, Ruan2014, Amari2014nat};
moreover, direct observations of pre-existence of MFR have also been
reported frequently in the SDO era with the help of high-quality
multi-wavelength data of the AIA \citep{Cheng2012, ZhangJ2012,
  LiL2013, LiT2013, LiT2013a, Patsourakos2013, YangS2014,
  ChenH2014}. Furthermore, the ideal MHD-instability theory has been
extensively applied in the investigations of filament eruptions
\citep{Romano2003, Williams2005, Green2007, Schrijver2008, LiuY2008,
  GuoY2010, ChengX2011, Nindos2012, XuY2012, LiuR2012, ChengX2013,
  Kliem2013, Jiang2014formation, Amari2014nat}, and it is noteworthy
that many successful and failed eruption events are in good consistent
with the theory, especially the theoretical threshold values for
instability are matched strikingly well in the
data-constrained/data-driven modeling results
\citep[e.g.,][]{Kliem2013, GuoY2010, Jiang2014formation,
  Jiang2013MHD}.

In this paper we provide further evidence for the pre-existence of MFR
and the ideal MHD-instability mechanism for the eruption of MFR by a
comparison study of an eruptive filament in AR~11283 and a nearby
non-eruptive filament. Both filaments have been studied independently
in our previous work \citep{Jiang2014formation, Jiang2014NLFFF} using
coronal field reconstructions, while here by comparing them with each
other, we provide further insight in the different nature between the
eruptive and non-eruptive filaments. We find that both of the
filaments are supported by coronal MFR. The MFR corresponding to the
AR eruptive filament is much smaller compared with the non-eruptive
MFR, but contains significantly more magnetic free energy than the
non-eruptive one. Both MFRs are weakly twisted and cannot trigger kink
instability. The eruptive MFR is unstable because its axis reaches
above a critical height for torus instability, while the non-eruptive
MFR is very firmly held by its overlying field, as its axis apex is
far below the torus-instability threshold height.

\begin{figure}[htbp]
  \centering
  \includegraphics[width=\textwidth]{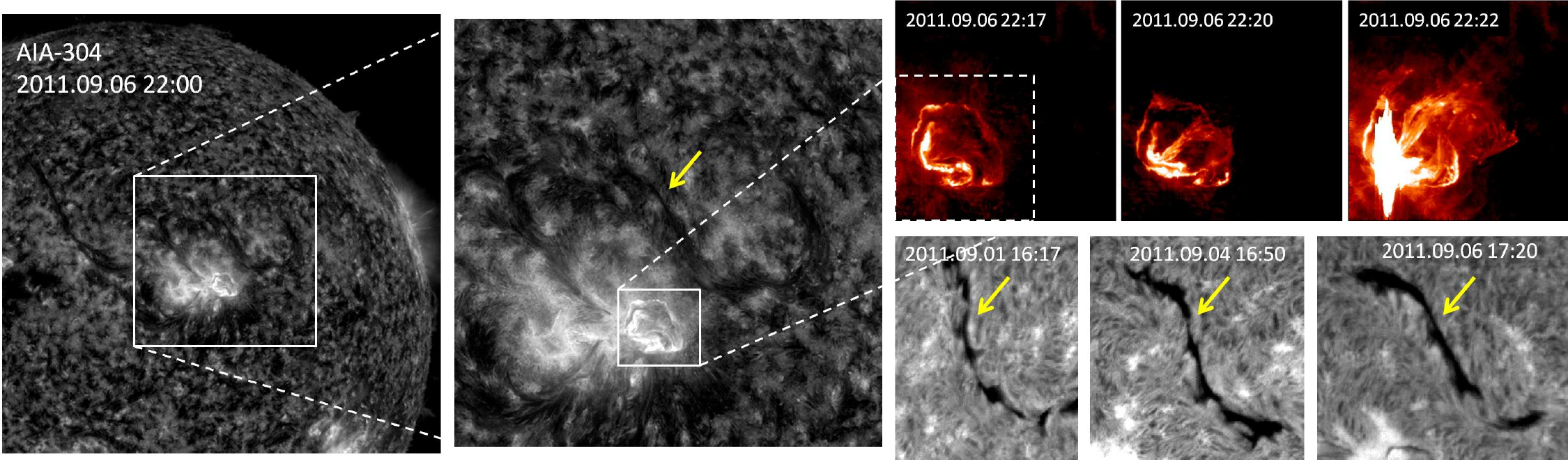}
  \caption{Left: a large AIA-304 image at 22:00~UT on September 6,
    showing the location of AR~11283. Middle: closer view of the
    region. The top-right small panels show the eruptive process of
    the AR filaments with three time snapshots, and bottom-right small
    panels show the BBSO-H$\alpha$ observations of the non-eruptive,
    large filament, with the arrows point to the filament main
    body as denoted in the AIA-304 image.}
  \label{fig:1}
\end{figure}

\section{Observations and Coronal Magnetic Field Model}
\label{sec:2}

AR~11283 produced two X-class flares (an X2.1 on September 6, 2011 and
an X1.8 on September 7), which have motivated many studies
\citep{WangS2012a, Petrie2012, Zharkov2013, FengL2013, Jiang2013MHD,
  Jiang2014formation, Ruan2014, LiuC2014, ShenY2014, XuY2014,
  ZhangQ2015}. Here we investigate the coronal configuration on
September 6 before the X2.1 flare. The SDO/AIA observed that an
S-shaped small-scale filament (with length of about 20~Mm) formed in
the core of AR and erupted at the onset of the flare at 22:12~UT, with
impulsive rising of the filament matter, as shown in
\Fig~\ref{fig:1}. On the other hand, in the northwest of the AR, there
is a large filament, the main body of which seen in H$\alpha$ shows a
slightly inverse-S shape with a length up to 200~Mm. The large
filament exists stably for many days, and survives during the eruption
of the small one, although its main body is very close to the flare
site. The right-bearing filament barbs as seen in the H$\alpha$
indicate that the non-eruptive filament should be related with a
left-handed twisted MFR, which is contrary to the eruptive one.

To investigate the coronal magnetic field that supports the filaments,
we have modeled the pre-flare magnetic field using our
CESE--MHD--NLFFF code with input of SDO/HMI vector magnetogram
\citep[see also][]{Jiang2014NLFFF}. The code is based on the MHD
relaxation approach and implemented by an advanced
conservation-element/solution-element (CESE) space-time scheme
\citep{Jiang2010}. We solve a set of zero-beta simplified MHD
equations with a fictitious frictional force to control the relaxation
process. The code is in line with our 3D MHD Data-driven Active Region
Evolution model \citep{Wu2006, Jiang2012c, Jiang2013MHD, Wu2013},
which has been well tested for reproducing both the slow quasi-static
evolution and extremely fast eruption process in the large-scale
corona. The present model is performed in a Cartesian box with the $z$
axis along the normal vector of the solar surface.

%++++++++++++++++++++++++++++++++++++++++++++++++++++++++++++

%\begin{figure}[htbp]
%  \centering
%  \includegraphics[width=\textwidth]{fig2}
%  \caption{Top: Comparison of the magnetic field lines with EUV
%    observations from SDO/AIA-171. The magnetic field lines in black
%    are plotted to compare with the bright coronal loops. The colored
%    rod-like lines represent a MFR along the large filament channel,
%    and the color denotes the height from the photosphere (in units of
%    Mm). The background images show the photospheric magnetic flux distribution.
%    Bottom: comparison of magnetic field lines and magnetic field
%    dips (the colored structure underlying the field lines) with
%    STEREO-A/EUVI-171 from the side view. The pink arrow marks the
%    filament and the yellow arrow marks its overlying arcades. (An
%    animation of the 3D field lines is provided).}
%  \label{fig:2}
%\end{figure}

\begin{figure}%[htbp]
  \centering
  \includegraphics[width=\textwidth]{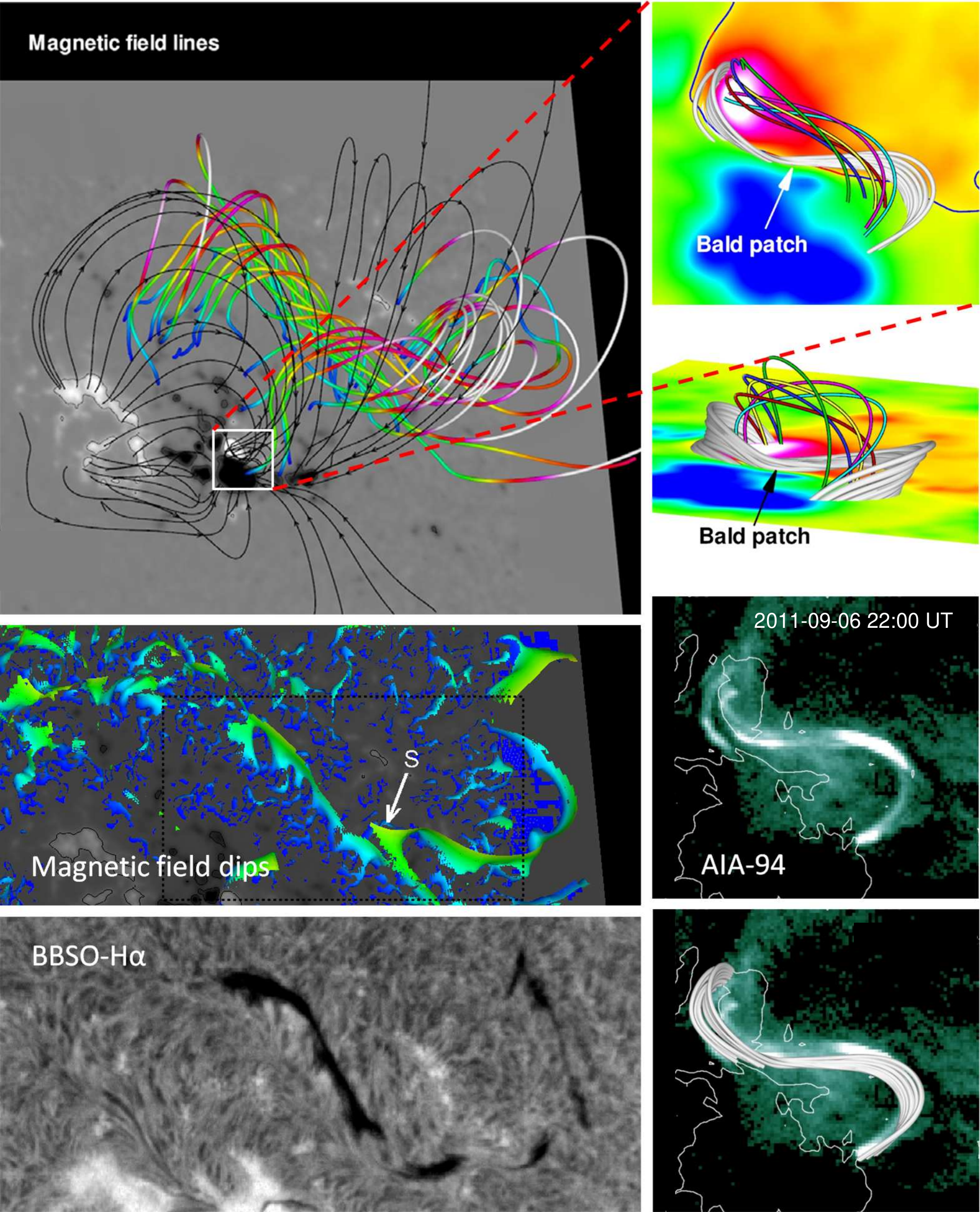}
  \caption{Left: Comparison of the magnetic dips of the large MFR with
    the large filament seen in BBSO H$\alpha$ (adapted from
    \citet{Jiang2014NLFFF}).  The top is the magnetic field lines, the
    middle is the magnetic dips (pseudo-colored by the height from 0
    to 60 Mm), and the bottom is the H$\alpha$ image. The dashed
      box denotes region of the main body of the filament.  The arrow
    'S' denotes a separation of the filament. Right: Comparison of the
    BPSS of the AR MFR with the AIA-94 sigmoid (adapted from
    \citet{Jiang2014formation}). The background color images show the
    photospheric magnetic flux distribution (white for $500$~G and
    blue for $-500$~G), with the PIL denoted by the curved line. The
    BPSS consists of the field lines plotted as white thick rods,
    which graze the photospheric surface at the BP. The colored field
    lines are sampled near the MFR axis. The bottom are the
      AIA-94 image of the sigmoid with the same field-of-view as the
      top ones. In the last panel the AIA image is overlaid by the
      BPSS field lines.}
  \label{fig:2}
\end{figure}

\section{Results and Analysis}
\label{sec:3}

\subsection{Comparison of coronal magnetic field model with observations}

%\Fig~\ref{fig:2} shows the magnetic field lines compared with the
%SDO/AIA-171 image and a limb view from STEREO-A/EUVI-171. A good
%agreement of the field lines with the coronal loops can be seen.

The coronal field model presents two MFRs of opposite helicity in the
same extrapolation box: a small one of right-hand twist (see a
close-up view in \Fig~\ref{fig:2}) corresponding to the eruptive AR
filament, and a large one of left-hand twist corresponding to the
non-eruptive intermediate filament.
%In the AIA-171 channel, the large filament
%channel is well seen as a much darker region among the bright loops,
%and the large MFR is co-spatial with the channel \citep[see Fig~3 of][]{Jiang2014NLFFF}.
%This filament was
%also observed by STEREO-A at the limb as a low-lying prominence, above
%which is a group of closed arcades (marked by a yellow arrow in
%\Fig~\ref{fig:2}). These arcades correspond to the potential-like
%field lines overlying the MFR, two ends of which are visible as bright
%rays in AIA-171, and they play the role of stabilizing the large MFR.
%On the other hand, the AR filament is not observed since it is
%fully embedded in a small core region that is behind the bright
%loops.
To validate the model result, we carried out a detailed comparison of
the magnetic field and its derived features with the observations of
the filament related structures. In particular, for the large-scale
filament, the H$\alpha$ observation is compared with the magnetic dips
in its MFR, since there is a good H$\alpha$ image of the filament
spine and barbs and these structures are believed to consist of cold
matter collected in the magnetic dips. While for the small AR
filament, since H$\alpha$ has no adequate resolution of it, comparison
of its MFR field lines with a EUV sigmoid that is closely co-spatial
with the AR filament provides a better way for our validation of the
modeling.

In the left panels of \Fig~\ref{fig:2}, we compare the magnetic dips
with the large H$\alpha$ filament. The magnetic dips are visualized by
showing part of the iso-surface ($B_z=0$) with $\vec B\cdot \nabla B_z
> 0$ (i.e., locations where the field lines are concave up). The dips
are also pseudo-colored by the height value for a better inspection of
the long extended dip structure, since there is a large amount of
localized fragmentation of dips (shown in deep blue) that is very
close to and on the photosphere (i.e., bald patches). Evidently, the
long extended dip reaches above 30~Mm, exhibiting an inverse S-shaped,
reproducing the main body of the filament. Especially, there are
small-scale magnetic dips emanating from the spine, and they match the
filament barbs strikingly well. The results are in line with and
provide an evidence for the MFR-dip model for filaments and their
barbs.

The right panels of \Fig~\ref{fig:2} show the comparison of the
magnetic field lines of the AR MFR with the sigmoid observed in AIA-94
(temperature of 6.3~MK). The sigmoid has a thin and enhanced forward-S
shape, indicating a right-hand twist. We find that this MFR has a bald
patch separatrix surface (BPSS), as shown by the thick white lines in
the figure, which is the MFR's outmost surface that touches the
photosphere. The BP is a part of the PIL on the photosphere where the
transverse field crosses from the negative \add{to} the positive
polarity (note that the BP is directly derived from the vector
magnetogram), and the BPSS consists of all the field lines that pass
through the BP. Theoretical study suggests that strong current sheet
can form in the BPSS of MFR and produces enhanced heating along the
surface, which manifests itself as the X-ray or EUV sigmoid
\citep{Titov1999}. The most important result is that the observed
sigmoid coincides with the BPSS field lines, but not with those near
the rope axis (the thick colored lines in the figure). Such fact not
only supports the BPSS current sheet model for sigmoid brightening,
but also confirms the validation of our model.

\begin{table}[htbp]
  \caption{Comparison of magnetic energy contents for the two MFRs.}
  \centering
  \begin{tabular}{lll}
    \hline
                             & AR Eruptive MFR & Non-eruptive MFR \\
    \hline
    Total energy             & $3.67 \times 10^{32}$~erg & $1.52 \times 10^{32}$~erg \\
    Potential energy         & $2.63 \times 10^{32}$~erg & $1.28 \times 10^{32}$~erg \\
    Free energy              & $1.04 \times 10^{32}$~erg & $0.24 \times 10^{32}$~erg \\
    Ratio of Free energy to potential energy & $40\%$   & $19\%$ \\
    Field volume             & $4.66 \times 10^{4}$~Mm$^{3}$  & $2.98 \times 10^{6}$~Mm$^{3}$ \\
    Mean free energy density & $2230$~erg~cm$^{-3}$      & $8$~erg~cm$^{-3}$ \\
    \hline
  \end{tabular}
  \label{tab:1}
\end{table}

\begin{figure}[htbp]
  \centering
  \includegraphics[width=\textwidth]{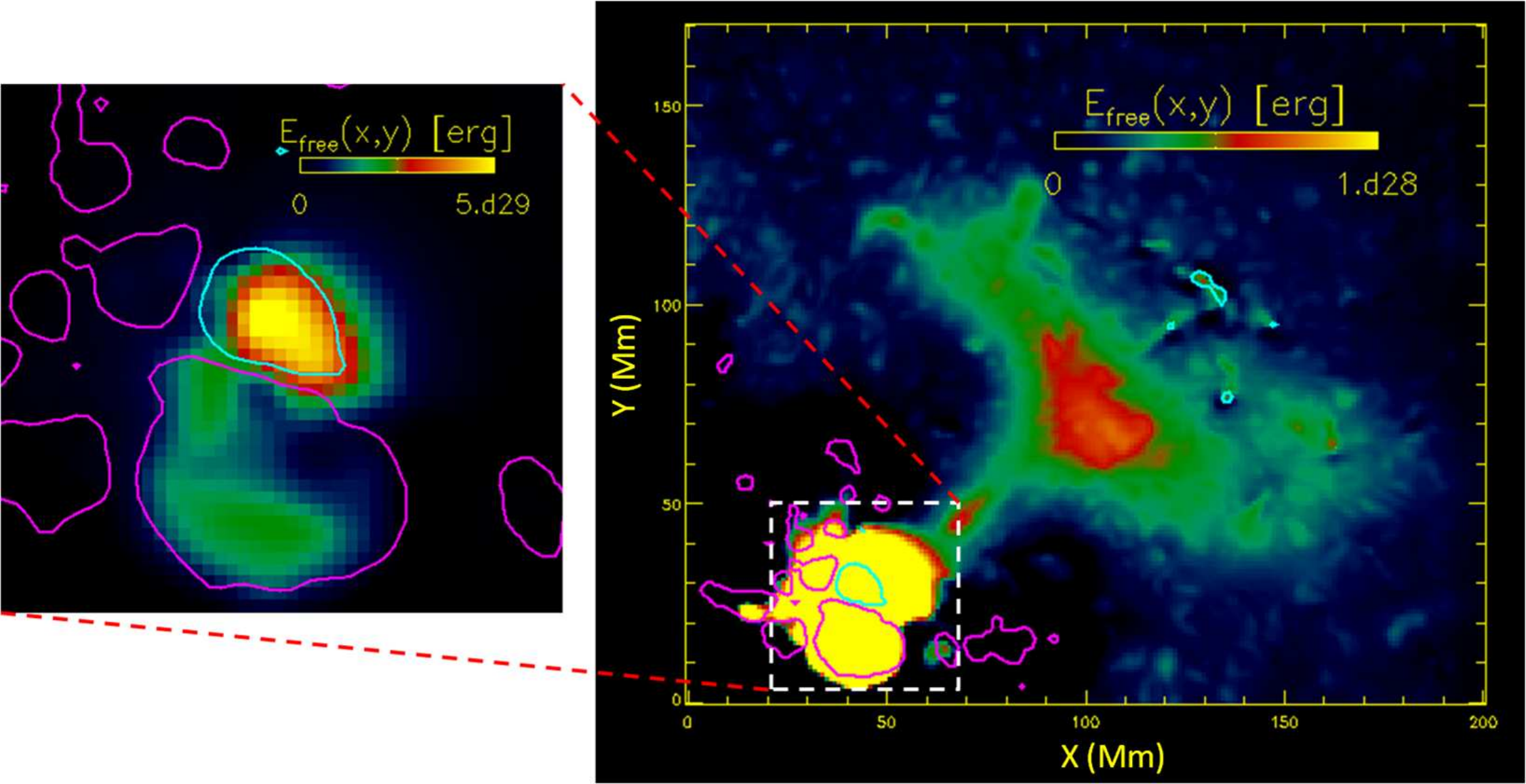}
  \caption{Free energy density integrated along the $z$ axis. The
    contour lines represent $B_{z}$ of $-500$~G (pink) and $500$~G
    (cyan). The small box in the large field of view shown in the
    right panel is denoted for the AR core, i.e., for the separated
    volume of the eruptive MFR, which is enlarged in the left
    panel. Since the energy is much stronger in the AR core than in
    intermediate region, different scales of the energy distribution
    are used in the two panels.}
  \label{fig:3}
\end{figure}

\subsection{Magnetic energy distributions}

In \Fig~\ref{fig:3} and Table~1 we compare the magnetic energy
distributions of the two filament-related MFRs. \Fig~\ref{fig:3} shows
the free energy density integrated along the $z$ axis, namely
\begin{equation}
E_{\rm free}(x,y) = dx dy \int \frac{B^2-B_{\rm pot}^2}{8\pi} dz .
\end{equation}
As can be seen, compared with the intermediate region, the free energy
is significantly stronger in the sunspot regions, i.e., the AR core,
because of the strong magnetic flux there. For the large MFR, there is
also clearly a channel of enhanced free energy along the corresponding
filament, with the central section possessing the strongest. Table~1
gives a comparison of different energy contents and mean free-energy
density of the two MFRs. While the two MFRs are involved in a whole
magnetic configuration, they can be regarded as independent of each
other according to the magnetic topology. As shown by
\citet{Jiang2014formation}, the AR eruptive MFR is formed with the
emerging of a new positive sunspot into the large negative flux
region, and a magnetic null-point fan separatrix surface (like a dome)
exists between the pre-existing system and the new emerged one. The AR
MFR with its overlying flux system is localized below the fan surface
with a height of about 30~Mm. As such, we roughly separate the two
MFRs in the full extrapolation volume using a small box with height of
30~Mm for the AR eruptive one as denoted in \Fig~\ref{fig:3}, and the
rest volume for the non-eruptive one. Then the different energy
contents are computed using these separated volumes for the two
MFRs. Notably, although in spatial size the eruptive MFR is much
smaller than the non-eruptive one, it contains most of the magnetic
free energy. Consequently the free energy density of the eruptive MFR
is far higher than that of the non-eruptive one. The ratios of free
energy to potential energy shows that the non-potentiality of the AR
MFR field is also significantly higher than that of the non-eruptive
one. These comparisons indicates a higher possibility of disruption of
the AR MFR than the other one.

\begin{figure}[htbp]
  \centering
  \includegraphics[width=\textwidth]{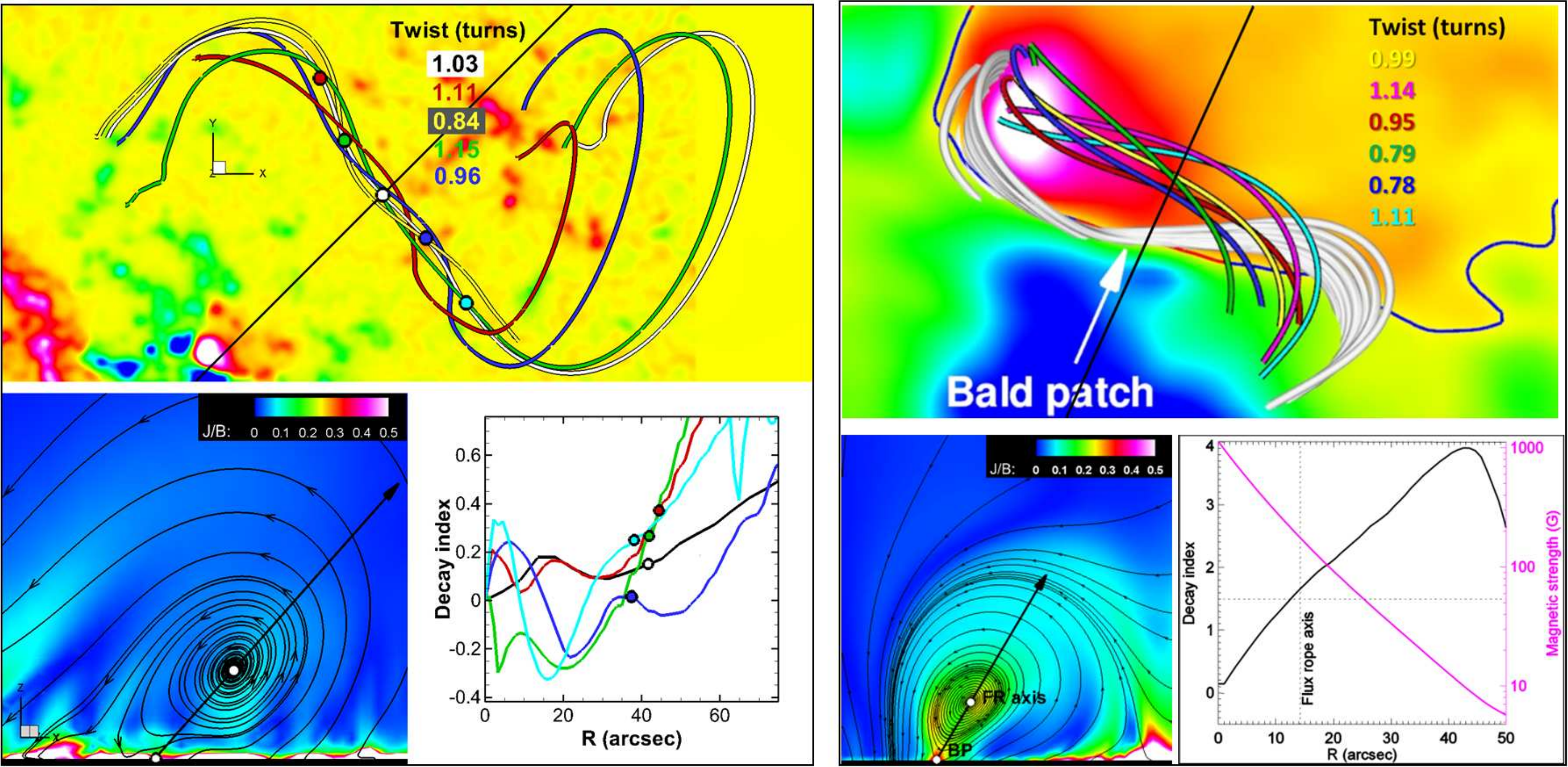}
  \caption{Analysis of the stability of the two MFRs: left panels for
    the non-eruptive one (adapted from \citet{Jiang2014NLFFF}) and
    right panels for the eruptive one (adapted from
    \citet{Jiang2014formation}). The top panels show the twist degrees
    of the MFRs, which are calculated for several sampled field lines
    around the rope axis (which is represented by the white line for
    the large MFR in the left panel and yellow line for the small MFR
    in the right panel, respectively). The bottom panels show the
    results for decay index. The central cross sections for the two
    MFRs are shown respectively, and these cross sections are
    vertically sliced along the black lines shown in the top
    panels. The streamlines in these cross sections show the 2D
    field-line tracing on the slices, which forms helical lines
    centered at the axis of the MFR. The arrowed thick lines started
    from the bottom and directed through the helix center (i.e., the
    rope axis) represent the paths along which the decay index is
    computed. Since the axis of the large MFR is very long, the
      decay index is calculated at four more points along its axis,
      which are denoted by the colored circles in the left panels.}
  \label{fig:4}
\end{figure}

\subsection{Stabilities of the Flux Ropes}

To further explain why the AR filament erupts immediately whereas the large
filament can keep stable, we study the stabilities of the MFRs with
the 3D coronal field in the context of ideal MHD
instabilities. Theoretical models and simulations suggest that a
coronal MFR (i.e., MFR with footpoints line-tied on the photosphere)
confined by an overlying potential arcade is subject to two kinds of
ideal instabilities, i.e., kink instability and torus
instability. Kink instability occurs if the twist degree of the MFR,
denoted by $T_n$, which measures the number of windings of the field
lines around the rope axis, exceeds a critical value. This threshold
value is estimated to be roughly $1.5 \sim 2$, according to several
studies \citep{Fan2003, Torok2004, Torok2005}. The torus instability
occurs when the outward expansion of a MFR due to its ``hoop force''
can no longer be confined by the overlying field (also referred to as
the external field) if the external field decreases sufficiently
fast. A decay index is defined as
\begin{equation}
n(R)=-\frac{\partial ({\rm ln} B_e)}{\partial({\rm ln} R)}
\end{equation}
to characterize how fast the external field $B_e$ decreases with
distance $R$ from the photosphere, and it has been found that if the
apex of the rope axis reaches a location with the decay index greater
than a threshold $n_c$, the MFR system is torus unstable.  This value
$n_c$ is expected to lie in the range of $1.1 \lesssim n_c \lesssim 2$
from a series of investigations \citep{Bateman1978, Kliem2006,
  Torok2007, Fan2007, Schrijver2008, Demoulin2010, Aulanier2010}.  One
should bear in mind that these threshold values for instabilities may
change depending on many factors, e.g., the specific shape of the flux
rope and the magnetic environments. Thus they should be used with
cautions in the realistic cases.

The twist degree $T_n$ of a force-free MFR can be simply quantified by
integral the force-free factor $\alpha = \vec J\cdot \vec B/B^2$ along
a given field line, namely,
\begin{equation}
T_n = \frac{1}{4\pi} \int \alpha dl
\end{equation}
\citep{Berger2006, Inoue2011, Inoue2013}. In the upper panels of
\Fig~\ref{fig:4}, we show the results for several field lines around
the rope axis. Since in the realistic case the MFR is actually a
bundle of field lines winding around each other, it is difficult to
locate precisely the axis of the rope like in an analytic or idealized
model.  We thus approximately find the rope axis by first making a
central cross section of the MFR, which is a vertical slice cutting
roughly through the middle of the MFR in a direction perpendicular to
the photospheric central PIL, and then assuming the rope axis as the
field line passing through the center of the helical shapes formed by
the poloidal flux of the rope (as shown in bottom panels of
\Fig~\ref{fig:4}). The results show that the twist degrees of both
MFRs are very close to one turn, although different field lines of
them have slightly different values of twist. Such value of twist is
below the thresholds of kink instability that have been
reported. Besides we have not observed a clear rotation or writhe of
the erupting AR filament as it rises, which would otherwise occur in
the eruption of kinked MFR. We concluded that neither of the filaments
can trigger kink instability.

Several points are needed to be clarified with respect to study the
torus instability based on computing the decay index. First, since it
is difficult to separate the external field from the total field in
the model, we use the potential field that corresponds to the same
magnetogram as an approximation of the external field, following
\citet{Fan2007} and \citet{Aulanier2010}. Secondly, as mentioned, the
axis apex of the MFR is hard to precisely determined, and we thus
approximately locate it at the center of the helical shapes formed by
the poloidal flux of the rope in the central cross section. For the
large filament of which the axis is rather long,
we compute the decay index at four more points along its axis.
Third,
since the MFR and overlying field configuration is inclined with
respect to the radial (or vertical) direction, it will be more
reasonable to compute the decay index along a path approximately
following the inclination of the system than along the vertical
direction. Last, because the field parallel to the rising direction
actually does not contribute to the inward confining force, the decay
index is computed for only the perpendicular component of the
potential field \citep{ChengX2011, Nindos2012}. Accordingly, we
compute the decay index for the two MFRs and compare them in the
bottom panels of \Fig~\ref{fig:4}. As can be seen, the non-eruptive
MFR locates at heights with decay index smaller than $0.5$, which is far below the
threshold for the torus instability, thus it is very firmly held by
its overlying field. On the other hand, the AR MFR reaches slightly
above the height with decay index of $1.5$, which is the threshold of
torus instability found by \citet{Torok2007} and \citet{Aulanier2010}
using idealized numerical simulations. Thus our result supports the
threshold value as $1.5$, and suggests that the quick eruption of the
AR MFR is due to the torus instability.

\section{Conclusions}
\label{sec:4}

We have performed a comparative analysis of an eruptive filament in
AR~11283 and a nearby non-eruptive filament in the intermediate region
using a zero-beta MHD-relaxation model for near force-free
extrapolation \citep{Jiang2013NLFFF}. The extrapolated coronal
magnetic field based on the measured vector magnetogram given by
SDO/HMI shows that the two MFRs support the two filaments. Validation
of the modeling magnetic field is performed by comparing it with EUV
and H$\alpha$ observations, which shows that the eruptive MFR contains
a BPSS coinciding spatially with the pre-eruptive sigmoid, and the
non-eruptive MFR has magnetic dips matching the shape of the
non-eruptive filament. By comparison the MFRs with each other, the
following physical characteristics are found: (i) The eruptive
filament is only one tenth of the non-eruptive filament in size.  (ii)
The amount of free energy of the eruptive filament is five times that
of the non-eruptive one (iii) The mean free energy density of the
eruptive filament is more than two orders larger than the non-eruptive
one (iv) Both the MFRs are weakly twisted, thus could not 
%\del{be  triggering} 
trigger the kink instability (v) Evaluating the
decay index shows that the axis of the eruptive MFR (filament) reaches
above a critical height for torus instability, where the non-eruptive
MFR is firmly held by is overlying field, as its axis apex is far
below the threshold height. We suggest the measured mean free-energy
density may be a good characteristic indicator for a filament eruption
in addition to other properties. In summary, the energy storage and
the trigger mechanism are both important to filament eruptions, and it
is supported that MFR can exist prior to eruption and the torus
instability can trigger MFR eruption.

%%% BIBLIOGRAPHY %%%%%%%%%%%%%%%%%%%%%%%%%%%%%%%%%%%%%%%%%%%%%%%%%%%%%%%%%%%
% \mbox{}~\\
% \noindent {\normalsize \bf Bibliography Included with \BibTeX }\\*
%       % more powerful
%   With \BibTeX\ the formatting will be done automatically for all
% the references cited with one
% of the \verb+\cite+ commands (Section~\ref{S-references}).
% Besides the usual items, it includes the title of the article
% and the concluding page number.

     % format of references provided by the journal (.bst)

\normalem
\begin{acknowledgements}

  This work is supported by the 973 program under grant
  2012CB825601, the Chinese Academy of Sciences (KZZD-EW-01-4), the
  National Natural Science Foundation of China (41204126, 41231068,
  41274192, 41031066, and 41374176), and the Specialized Research Fund
  for State Key Laboratories. C.W.J., S.T.W. and Q.H. are also supported by NSF-AGS1153323 and AGS1062050.
   Data are courtesy of NASA/{SDO} and the
  HMI science teams. We appreciate helpful discussions with
  T.~T{\"o}r{\"o}k and B.~Kliem.

\end{acknowledgements}

\newpage

%\bibliographystyle{raa}
%\bibliography{all}

\end{document}